%% file: ETFA_23.tex
\documentclass[twocolumn,showpacs,%
nofootinbib,aps,superscriptaddress,%
eqsecnum,prd,notitlepage,showkeys,10pt]{revtex4-1}
\usepackage{amsmath,amssymb,amsfonts}
\usepackage{graphicx}
\usepackage{textcomp}
\usepackage{xcolor}

\usepackage[noend]{algpseudocode}
\usepackage{algorithm}
\usepackage{comment}
\usepackage[hidelinks]{hyperref}
\makeatletter 
\newcommand{\linebreakand}{%
\end{@IEEEauthorhalign}
\hfill\mbox{}\par
\mbox{}\hfill\begin{@IEEEauthorhalign}
}
\makeatother 

\def\BibTeX{{\rm B\kern-.05em{\sc i\kern-.025em b}\kern-.08em
		T\kern-.1667em\lower.7ex\hbox{E}\kern-.125emX}}
\begin{document}
	
	\title{
		Structural Analysis of GRAFCET Control Specifications
	}

	\author{Aron Schnakenbeck}
	\affiliation{
		\textit{Institut für Automatisierungstechnik} \\
		\textit{Helmut-Schmidt-Universität}\\
		22043 Hamburg, Germany \\
		\{aron.schnakenbeck, alexander.fay\}@hsu-hh.de
	}
	\author{Robin Mroß}
	\affiliation{
		\textit{Lehrstuhl Informatik 11} \\
		\textit{RWTH Aachen University}\\
		52074 Aachen, Germany \\
		\{mross, voelker, kowalewski\}@embedded.rwth-aachen.de
	}
	\author{Marcus Völker}
	\affiliation{
		\textit{Lehrstuhl Informatik 11} \\
		\textit{RWTH Aachen University}\\
		52074 Aachen, Germany \\
		\{mross, voelker, kowalewski\}@embedded.rwth-aachen.de
	}
	\author{Stefan Kowalewski}
	\affiliation{
		\textit{Lehrstuhl Informatik 11} \\
		\textit{RWTH Aachen University}\\
		52074 Aachen, Germany \\
		\{mross, voelker, kowalewski\}@embedded.rwth-aachen.de
	}
	\author{Alexander Fay}
	\affiliation{
		\textit{Institut für Automatisierungstechnik} \\
		\textit{Helmut-Schmidt-Universität}\\
		22043 Hamburg, Germany \\
		\{aron.schnakenbeck, alexander.fay\}@hsu-hh.de
	}
		
	\begin{abstract}
	The graphical modeling language GRAFCET is used as a formal specification language in industrial control design. 
	This paper proposes a structural analysis that approximates the variable values of GRAFCET to allow verification on specification level. 
	GRAFCET has different elements resulting in concurrent behavior, which in general results in a large state space for analyses like model checking.
	The proposed analysis approach approximates that state space and takes into consideration the entire set of GRAFCET elements leading to concurrent behavior. The analysis consists of two parts:
	We present an algorithm analyzing concurrent steps to approximate the step variables and we adapt analysis means from the field of Petri nets to approximate internal and output variables.
	The proposed approach is evaluated using an industrial-sized example to demonstrate that the analysis is capable of verifying behavioral errors and is not limited by the specification size of practical plants.
	\end{abstract}

	\maketitle
	%
	
	\input{p_introduction.tex}
	\input{p_related_work.tex}

	\input{p_preliminaries.tex}
	\input{p_contribution.tex}
	\input{p_evaluation.tex}
	\input{p_conclusion.tex}
	
	\begin{acknowledgments}
		This research is part of the project "Analysis of GRAFCET specifications to detect design flaws" (project number 445866207) funded by the Deutsche Forschungsgemeinschaft.
	\end{acknowledgments}
	
	\bibliographystyle{IEEEtran.bst}
	\bibliography{bibliography}
	
\end{document}

%% file: p_introduction.tex
\section{Introduction}
In industrial automation, Programmable Logic Controllers (PLC) are widely used. To design the control code running on a PLC, a beneficial approach is to use formal means in order to first specify the logical behavior of the PLC before implementing the control code \cite{Schumacher.13b}. 
Using a formal specification in the design phase has multiple advantages like using the specification as documentation and communication tool, allowing an automatic transformation into control code and applying formal verification at specification level \cite{Julius.19}. 

In this context, IEC 60848 GRAFCET \cite{iec60848} can be used as a description means. 
GRAFCET is a graphical, semi-formal, domain-specific language to model, e.g., control code of PLCs.
GRAFCET is used in several industrial domains like railway transport and the manufacturing industry and is widely known in the respective areas \cite{Provost.11}. This acceptance of GRAFCET might improve the acceptance of formal methods in the respective domains, which is still a problem \cite{VogelHeuser.14}.
Although GRAFCET adapts concepts of Petri nets -~like transitions and steps, connected alternately by arcs~- it provides a considerable number of additional modeling mechanisms like hierarchical structuring of the specification which allow for compact modeling of complex systems \cite{Mross.22}. 
Regarding the application of formal methods to GRAFCET specifications, there is preliminary work by Julius et al. \cite{Julius.17} to allow a code generation of such hierarchical GRAFCET specifications to PLC-code.
Because the work presented by Julius et al. does not cover verification of the Grafcets (the term \textit{Grafcet} refers to an instance of GRAFCET), we extend the approach with formal verification.
A verification on specification level has the advantage of finding possible design errors early in the design process, given that the costs of correcting errors in software systems increases exponentially as the development phase progresses \cite{Boehm.1981}.

The verification approach proposed in this work is a structural analysis that approximates the variable values of GRAFCET and takes elements into account proposed by the standard that result in concurrent behavior. 
The term \textit{structural analysis} originally refers to a type of verification of Petri nets that uses algebraic tools that do not require to build the reachability graph \cite{Cabasino.13}, resulting in a linear relaxation of the reachability graph. Similarly, the structural analysis of GRAFCET proposed in this work approximates the GRAFCET behavior using, among others, a linear relaxation. Further, the structural analysis of GRAFCET is performed on the basis of its structure, i.e., how the steps and transitions are connected by arcs. It does not take into account its transition conditions and, therefore, not the internal variables. Hence, structural analysis is a subtype of static analysis, which in general approximates the states of a program and therefore, its behavior without executing it \cite{Praehofer.12}.
The structural analysis proposed in this work is able to approximate the reachability and concurrency of steps as well as the values of output variables. This information can be used to identify safety critical situations or possible race conditions. The resulting over-approximation leads to the fact that false alarms can take place.

We will compare the proposed approach to other possible approaches in Sec.~\ref{sec:relWork}, followed by the preliminaries in Sec.~\ref{sec:prelim} on GRAFCET and on analysis means from the field of Petri nets that are applied to GRAFCET in Sec.~\ref{ssec:flowinsens}.  
In Sec.~\ref{sec:contribution} we point out why concurrent structures are challenging (Sec.~\ref{ssec:problem}) and present an approach that is twofold: In Sec.~\ref{ssec:stepApprox} we present an algorithm that calculates reachable and concurrent steps and, therefore, approximates the values of step variables. In Sec.~\ref{ssec:flowinsens} we present an analysis using means from the field of Petri nets to obtain information about how often actions can be executed to approximate the values of internal and output variables.
We end with evaluating the proposed analysis on a practical example (Sec.~\ref{sec:eval}) before giving a conclusion (Sec.~\ref{sec:conclusion}).

%% file: p_related_work.tex
\section{Related work}
\label{sec:relWork}

For verifying GRAFCET there are approaches suitable for model checking, such as translating hierarchical Grafcets into time Petri nets by Sogbohossou et al. \cite{Sogbohossou.20} and recently transforming Grafcets into Guarded Action Language by Mroß et al. \cite{Mross.22}. 
Utilizing a model checking approach allows for an exhaustive exploration of the model but has the disadvantage of being limited by state space explosion for practical application.

Very few approaches are presented for analyzing GRAFCET without applying model checking.
A structural analysis regarding the hierarchical dependencies between modules of the Grafcets (called \textit{partial Grafcets}) is presented by Lesage et al. \cite{Lesage.93}. The authors provide an analysis to ensure that the hierarchical dependencies form a partial order. Moreover, Lesage et al. \cite{Lesage.96} provide an analysis of the GRAFCET-specific expressions by extending the Boolean algebra by events represented by rising and falling edges of Boolean signals in GRAFCET. This allows the user to check syntactic properties of transition conditions.
Both presented approaches \cite{Lesage.93, Lesage.96} are only capable of detecting syntactical design flaws regarding the structure of the Grafcets and not design flaws regarding the behavior of the Grafcets like the reachability of safety critical situations.
Schumacher et al. \cite{Schumacher.13a} present an approach to transform the time constraints of GRAFCET into Control Interpreted Petri nets (CIPN), a specific kind of Petri net. They later extended the approach in \cite{Schumacher.14} to normalize hierarchical Grafcets and formalize them as CIPN. The verification of the formalized control specifications is not covered by Schumacher et al.
The approach, in fact, forms the groundwork for structural analyses based on methods known from the field of Petri nets. Well established analysis tools such as those described in \cite{Bonet.07} already exist for the structural analysis of Petri nets and we present how to apply them to partial Grafcets in Sec.~\ref{ssec:pn}, since the structural analysis known from the fields of Petri nets have to be adapted for the peculiarities of GRAFCET.

The normalization technique proposed in \cite{Schumacher.14} is not adopted in this work. It replaces the implicit, hierarchical flow relations between partial Grafcets induced by enclosing steps and forcing orders (cf. Sec. \ref{ssec:syntax}) by transitions and steps. Therefore, the Grafcets' hierarchical information is lost during the process.
Hierarchical information, however, is valuable because it can be exploited during the verification process in terms of a modular analysis of the partial Grafcets.

All cited approaches are limited by state space explosion and none of them allows for a verification of behavioral design flaws which would occur during run-time.
Therefore, this work aims to present a structural analysis that does not suffer from a state space explosion and is able to detect behavioral errors.

%% file: p_preliminaries.tex
\section{Preliminaries}
\label{sec:prelim}
Preliminaries on GRAFCET syntax and analytical methods based on linear algebra are explained in this section.

\subsection{Syntax of IEC 60848 GRAFCET}
\label{ssec:syntax}
Since the GRAFCET standard does not define the syntax of GRAFCET sufficiently for formal verification, we use the formalization proposed by Mroß et al. \cite{Mross.22} in this work to explain the concepts of GRAFCET that are important for this contribution.

A Grafcet $G = (V_\mathit{in}, V_\mathit{int}, V_\mathit{out}, C)$ comprises a set of partial Grafcets $C \neq \emptyset$ with globally available sets of input variables $V_\mathit{in}$, internal variables $V_{int}$ and output variables $V_\mathit{out}$. Variables can either be Boolean or integral, i.e. $v$ is assigned a value of $\mathbb{Z}$ for all $v \in V_\mathit{in} \cup V_\mathit{int} \cup V_\mathit{out}$ with Boolean variables being limited to the set $\{0, 1\}$. 
Given these variables, we can construct Boolean expressions with usual relational symbols (such as $=$ and $\le$) and Boolean operators (such as disjunction $\lor$ and negation $\lnot$). A variable may change values caused by an event. 
By $\mathit{CND}$ we denote the set of all Boolean expressions over variables in $G$. 
Every partial Grafcet $c \in C$ is a 6-tuple $c = (S, I, E, M, T, A)$, where 
\begin{itemize}
	\item $S$ is a finite set of steps, each of which is either active or inactive,
	\item $I \subseteq S$ is the set of initial steps,
	\item $E \subseteq S \times C$ is the set of enclosing steps,
	\item $M \subseteq S$ is the set of marked steps,
	\item $T \subseteq \mathcal{P}(S) \times \mathcal{P}(S) \times \mathit{CND}$ is the set of transitions, where $\mathcal{P}$ denotes the power set, and
	\item $A$ is a set of actions.
\end{itemize}
\begin{figure}[t]
	\centering
	\includegraphics[width=.6\columnwidth]{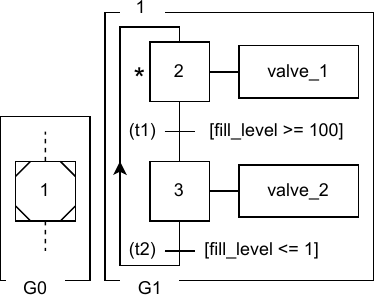}
	\caption{Illustrative example of a Grafcet.}
	\label{fig:exampleGrafcet}
\end{figure}
We use the notation $S_{c}$, $I_{c}$, $E_{c}$, $M_{c}$, $T_{c}$, $A_{c}$ to refer to the respective sets of a given partial Grafcet $c \in C$. The set $M_{c}$ describes the steps that are activated by the enclosing step. Every $e \in E_{c}$ describes an enclosing step, which translates formally to $e = (s, c_\mathit{enc})$ for a $s \in S_{c}$ and a partial Grafcet $c_\mathit{enc} \in C$. If an enclosing step becomes active, it activates all steps $m \in M_{c_\mathit{enc}}$. If an enclosing step becomes inactive, it deactivates all steps $s \in S_{c_\mathit{enc}}$. We say that $c$ is \textit{enclosed} iff $M_{c} \neq \emptyset$. 
Every step $s \in S_{c}$ induces a new Boolean variable $x_{s}$ which indicates the activation status of $s$ and is true iff the step is active in the current situation. These variables can be used in Boolean expressions $\mathit{CND}$. 
Fig.~\ref{fig:exampleGrafcet} shows an illustrative example of a Grafcet consisting of two partial Grafcets $C = \{\mathrm{G0, G1}\}$. G1 has two steps $S_{\mathrm{G1}} = \{2, 3\}$ one of which is an enclosing step $ M_{\mathrm{G1}} = \{2\}$ and two transitions $T_{\mathrm{G1}} = \{\mathrm{t1}, \mathrm{t2}\}$ as well as two continuous actions associated to step~2 and step~3. G1 is enclosed by step~1, indicated by the 1 at the top of G1, and therefore $E_{\mathrm{G0}}=\{(1, \mathrm{G1})\}$. If step~1 is activated, step~2 is activated as well and G1 can evolve freely as long as step~1 stays active.

A transition $t \in T_{c}$ is a triple $t = (\bullet t, t \bullet, b)$, where 
$\bullet t \subseteq S_{c}$ is the set of immediately preceding steps, 
$t \bullet \subseteq S_{c}$ is the set of immediately succeeding steps, 
$\bullet t \neq \emptyset \lor t \bullet \neq \emptyset$
and $b \in \mathit{CND}$ is the transition condition.
We also call $\bullet t$ the \textit{upstream} and $t \bullet$ the \textit{downstream} of $t$.
We say that $t$ is \textit{enabled} if $x_{s}$ is true for every $s \in \bullet t$. We say that $t$ can \textit{fire} if it is enabled and $b$ is true. Similarly to the upstream and downstream of transitions we define $\bullet s \subseteq T_c$ as the set of immediately preceding transitions of $s$ and $s\bullet \subseteq T_c$ as the set of immediately succeeding transitions of $s$. 

\begin{figure*}[t]
	\centering
	\includegraphics[width=.8\textwidth]{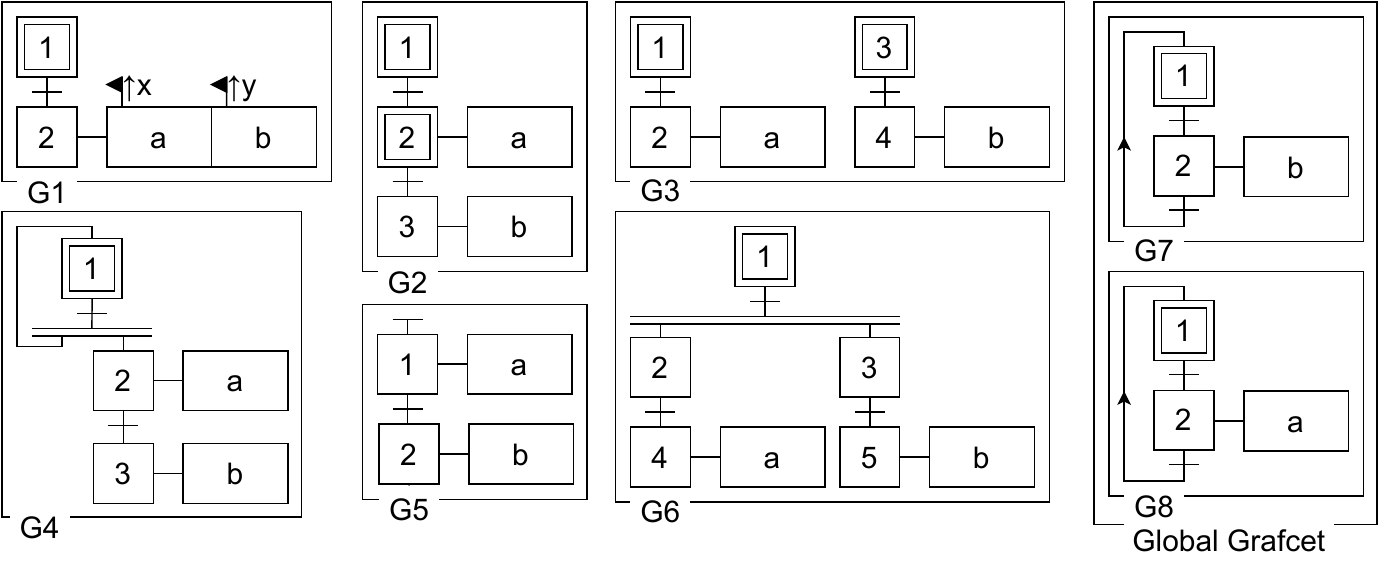}
	\caption{Different structures in GRAFCET \cite{iec60848} resulting in concurrent behavior indicated by actions $a$ and $b$.}
	\label{fig:concurrentStructures}
\end{figure*}

Finally, we formalize the set of actions $A_{c}$. The standard defines different types of actions: continuous actions ($A_{cont}$), stored actions ($A_\mathit{stor}$) and forcing orders ($A_\mathit{fo}$). 
These sets are assumed to be disjoint. Let $A_{c} = A_\mathit{cont} \cup A_\mathit{stor} \cup A_{\mathit{fo}}$. Every element of $A_\mathit{cont}$ is a triple $(s, v, b)$, where 
$s \in S_{c}$ is the associated step, 
$v \in V_\mathit{out}$ is an output variable which must be Boolean and
$b \in \mathit{CND}$ is the action condition.
We say that a continuous action is \textit{active} if $x_{s}$ and $b$ are true. Several partial Grafcets in $G$ may employ continuous actions on the same output variable $v$. In this case, $v$ is set to true if at least one of these continuous actions is active. Note that $v$ can not be used by any stored action.
Every element of $A_\mathit{stor}$ is a tuple $(s, v, val, b)$, where 
$s \in S_{c}$ is the associated step, 
$v \in V_\mathit{int} \cup V_\mathit{out}$ is an internal or output variable, 
$val$ is an expression yielding a value in the respective domain, e.g. $val \in \mathbb{Z}$ and 
$b \in \mathit{CND}$ is the action condition.
A stored action sets $v$ to $val$ if $x_{s}$ and $b$ are true. This also allows to model actions on activation and deactivation of a step, as introduced by the standard. 
Finally, every element of $A_\mathit{fo}$ is a tuple $(s, c_\mathit{forced}, S)$, where 
$s \in S_{c}$ is the associated step, 
$c_\mathit{forced} \in C$ is the partial Grafcet which is to be forced and 
$S \in (\mathcal{P}(S_{c_\mathit{forced}}) \cup \{*, \mathit{init}\})$.
A forcing order is regarded as a special kind of continuous action. It is active while $x_{s}$ is true and forces $c_\mathit{forced}$ into the situation specified by $S$. If $S = *$, then the current situation in $c_\mathit{forced}$ is retained for as long as $s$ is active. If $S = \mathit{init}$ then $c_\mathit{forced}$ is set to its initial situation. Otherwise, it is set to the specified situation (element of the power set $\mathcal{P}(S_{c_\mathit{forced}})$).

\subsection{Petri net analysis means adapted for GRAFCET}
\label{ssec:pn}

Schumacher et al. \cite{Schumacher.14} have shown how a Grafcet can be interpreted as a CIPN. We use this interpretation to adapt analysis means from the field of Petri nets to obtain structural information about each partial Grafcet.
In particular we want to adapt analysis means based on linear algebra, so-called S- and T-invariants which are described in more detail, e.g., in \cite{Girault.03}.

Analog to the formalization of a Petri net Schumacher et al. define an $|S|\times |T|$ incidence matrix $\mathbf{N}$ of a so-called basic Grafcet, where a basic Grafcet is a Grafcet without hierarchical elements. The elements $n_{ij}$ of $\mathbf{N}$ are defined as $n_{ij} = -1$ if $s_i \in \bullet t_j$, $n_{ij} = 1$ if $s_i \in t_j\bullet$ and $n_{ij} = 0$ otherwise. Neglecting the transition conditions, a linear relaxation of the dynamic behavior of the basic Grafcet can be described by the so-called state equation $\mathbf{sit} = \mathbf{sit}_0 + \mathbf{N} \mathbf{q}$, where $\mathbf{sit}$ is the situation of the Grafcet (i.e., a vector of the step variables), $\mathbf{sit}_0$ the initial situation and $\mathbf{q}$ is the firing count vector stating how often a transition fires until $\mathbf{sit}$ is reached from $\mathbf{sit}_0$.
The incidence matrix $\mathbf{N}$ can be analyzed using T- and S-invariants. A T-invariant is a vector $\mathbf{x}$ such that $\mathbf{N}\mathbf{x} = 0$. T-invariants can detect possible loops in the reachability graph of Grafcet 
since $\mathbf{N}\mathbf{x} = 0 = \mathbf{sit}' - \mathbf{sit}$, where $\mathbf{sit}'$ can be reached from $\mathbf{sit}$ when the transitions in $\mathbf{x}$ fire.
A S-invariant is a vector $\mathbf{y}$ such that $\mathbf{y}^T \mathbf{N} = 0$, where $T$ denotes to transposed. For Petri nets S-invariants indicate an upper bound for a possible number of tokens in a place, since $\mathbf{y}^T\, \mathbf{sit} = \mathbf{y}^T \,\mathbf{sit}_0 + \mathbf{y}^T\mathbf{N}\mathbf{q}$ $\Leftrightarrow$ $\mathbf{y}^T\, \mathbf{sit} = \mathbf{y}^T \,\mathbf{sit}_0$, where the firing vector $\mathbf{q}$ has no influence on the ratio of tokens. 
This is not directly applicable to GRAFCET since the steps induce a binary activity variable. However, S-invariants applied to GRAFCET indicate if the number of a step's activation in a Grafcet is bounded to a value $n \in \mathbb{N}$ and $n < |S|$, where $n$ is the maximum value in the S-invariants for the corresponding step. 

Both types of invariants are used in Sec.~\ref{ssec:flowinsens} to estimate how often an action can be executed and therefore, approximate the internal and output variables. 

%% file: p_contribution.tex
\section{Structural analysis of GRAFCET}
\label{sec:contribution}
A static analysis of GRAFCET based on its control flow without building the state space or part of it is challenging since concurrent actions can change the global variable values, but the concurrency is not directly visible from an arbitrary point in the control flow.
To approach this challenge, we first consider the different elements in GRAFCET resulting in concurrent behavior in Sec. \ref{ssec:problem}, before we present a structural analysis of GRAFCET in Sec.~\ref{ssec:stepApprox} that identifies concurrent steps along the control flow to make the concurrency visible at arbitrary points in the control flow.
In Sec.~\ref{ssec:flowinsens} we present an analysis that approximates the internal and output variable values taking the concurrent behavior into account.

\subsection{Problem definition for the structural analysis of GRAFCET}
\label{ssec:problem}

Before applying a control flow based analysis we need to define the control flow of GRAFCET.
In GRAFCET, instructions that read variable values are connected to conditions associated with transitions, 
and instructions that write variable values are connected to actions associated with steps. Therefore, the statements of the control flow correspond to steps and transitions which are connected by arcs forming the flow relations (in Sec.~\ref{ssec:syntax} formalized via $\bullet t$ and $t \bullet$).
This relation we want to use to analyze the behavior of the GRAFCET.
Besides the arcs, Grafcets can have additional flow relations induced by hierarchical elements like enclosing steps and forcing orders. Those relations are always between different partial Grafcets.

In comparison to sequential programs (i.e., programs written for example in C running on
a single thread) statements of the Grafcet's control flow can be executed concurrent to each other. This is a problem when the statements depend on each other and their execution order is non-deterministic. Particularly steps can be activated in parallel and their associated actions are executed concurrently in a non-deterministic order due to changes of input variables. 
An example of two such actions could be the execution of the value assignments $x := 0$ and $x := x +1$ where the execution order has an influence on the resulting value of $x$.
Therefore, analysis means from the field of sequential programs are not applicable to GRAFCET.

The GRAFCET standard \cite{iec60848} presents different structures resulting in concurrent behavior as shown in the partial Grafcets G1 to G8 in Fig. \ref{fig:concurrentStructures}:
\begin{itemize}
	\item Multiple conditional actions (graphically represented by a flag, followed by an expression like $\uparrow\! x$, where $\uparrow$ is called a rising edge of $x$ and occurs when $x$ changes from 0 to 1) associated to a single step (G1)
	\item Multiple initially active steps in a sequence (G2) or in parallel (G3)
	\item Elements producing active steps like source transitions ($\bullet t = \emptyset$ in G5) or its equivalence using an activation of parallel sequences (G4) as introduced by the standard \cite{iec60848}
	\item Activation of parallel sequences activating multiple steps at the same time ($|{t \bullet}| > 1$ in G6)
	\item Concurrently activated partial Grafcets (G7 and G8)   
\end{itemize}
All these structures can result in a non-deterministic firing order of transitions and a non-deterministic execution order of actions. The latter is indicated in Fig. \ref{fig:concurrentStructures} by actions $a$ and $b$ in concurrent parts of the Grafcet. 
Only the last structure containing G7 and G8 occurs in relation to a hierarchical structuring indicated by the Global Grafcet notation enclosing the partial Grafcets G7 and G8. 

Besides the fact that the order of firings and executions is non-deterministic, their number of executions (i.e., how often a transition or action is executed) is non-deterministic as well. 
E.g., source transitions can non-deterministically generate multiple active steps in a subsequent sequence due to the non-deterministic change of input variables. Structures like shown in G4 in Fig.~\ref{fig:concurrentStructures} have a similar behavior.\\

\subsection{Analysis of the reachability and concurrency of steps}
\label{ssec:stepApprox}
To analyze the presented behavior of GRAFCET, we propose an algorithm (in which $A \triangleleft B$ is short for $A \leftarrow A \cup B$) 
that approximates the reachability (Alg. \ref{alg:reachabilility}) and concurrency (Alg. \ref{alg:concurrency}) of steps. 
Both algorithms only take the structure of the Grafcet into account, i.e., assuming the transition conditions evaluate to true. 
Because of this approximation, the analysis is independent of variable values affected by potential race conditions which allows to take the concurrent elements of GRAFCET into account.

\begin{algorithm}[H]
	\caption{Analysis of reachable steps}
	\label{alg:reachabilility}
	\begin{algorithmic}[1]
		\Function{ReachAnalysis}{initial Worklist $W$, partial Grafcet $c \in C$, initially concurrent steps $\mathcal{S}$}
		\While {$\exists t \in W$}
		\State $W \leftarrow W\backslash \{t\}$
		\State $\mathcal{S'} \leftarrow \mathcal{S}$
		
		\If {$s' \in S^R$ holds for all $s' \in {\bullet t}$}
		\ForAll{$s \in t \bullet$}
		\If {$s \notin S^R$}
		\State $S^R \triangleleft \{s\} $
		\State $W \triangleleft {s \bullet}$
		\EndIf
		\State $\mathcal{S}\, \triangleleft\,$\textsc{ConcurrAnalysis}$(c,$ $\mathcal{S}, t, s)$
		\EndFor
		\EndIf
		\ForAll{$s'' \in S_c$}
		\If {${S^C_{s''}}'\in\mathcal{S'}  \not =S^C_{s''}\in\mathcal{S}$}
		\State $W \triangleleft s''\bullet$
		\EndIf
		\EndFor
		\EndWhile
		\State \textbf{return} $S^R, \mathcal{S}$
		\EndFunction
	\end{algorithmic}
\end{algorithm}

To analyze the reachable steps, we propose a worklist algorithm, presented in Alg. \ref{alg:reachabilility}. The algorithm  analyzes the flow along the transitions and calculates a set $S^R \subseteq S_c$ of reachable steps for a given partial Grafcet $c \in C$ and its initial situation $S^I \subseteq S_c$ which will be defined in a moment. 
For a transition $t$ from the worklist the algorithm examines if all upstream steps $s' \in \bullet t$ are reachable (line 5).  
This is an over-approximation of a transition being enabled. The downstream steps $ s \in t \bullet$ are marked as reachable and in turn their downstream transitions $s \bullet$ are put on the worklist until the analysis stabilizes. For every step $s$ that is reached this way we calculate its concurrent steps $S^C_s$ using \textsc{ConcurrAnalysis} which is discussed below. Since the sets of concurrent steps $\mathcal{S}$ can still change after the respective steps are marked as reachable, the stabilization has a second criteria: If the algorithm calculates additional concurrent steps $S_{s''}^C$ to a step $s''$ its downstream transitions are put on the worklist (line 11-13) to propagate them along the transitions as discussed below.

The initialization of $W$ covers the control flow caused by enclosing steps and forcing orders.
As presented in \cite{Lesage.93}, hierarchical dependencies form a partial order. Each of these dependencies yields an initial situation $S^I \subseteq S_c$ in the inferior partial Grafcet $c$. Therefore, every partial Grafcet might have multiple initial situations $S^I$: initial steps induce $S^I \leftarrow I_c$, for enclosings the initial situation are the steps activated by the enclosing step ($S^I \leftarrow M_c$) and for forcings the initial situation is the situation that is enforced by the forcing order ($S^I \leftarrow S_a$ which holds for all $a \in A_\mathit{fo}$ and $S_a \subseteq S_c$).
The worklist is initialized with the initially enabled transitions which are the downstream transitions $s\bullet$ of the initially active steps ($W \leftarrow \{s\!\bullet | s\in S^I\}$). 

\begin{algorithm}[H]
	\caption{Analysis of concurrent steps}
	\label{alg:concurrency}
	\begin{algorithmic}[1]
		\Function{ConcurrAnalysis}{partial Grafcet $c$, set of concurrent steps for every step $\mathcal{S}=\{S_s^C\}_{s \in S_c}$, current transition $t$, current step $s \in t \bullet$}
		\State $S_s^C \triangleleft {t\bullet} \backslash \{s\} $
		\State $S_s^C \triangleleft \bigcap_{s' \in \bullet t} S_{s'}^C$
		\ForAll {$s'' \in S^C_s$}
		\State $S_{s''}^C \triangleleft \{s\}$
		\EndFor		
		\State \textbf{return} $\mathcal{S}$
		\EndFunction
	\end{algorithmic}
\end{algorithm}
\begin{figure}[t]
	\centering
	\includegraphics[width=0.7\linewidth]{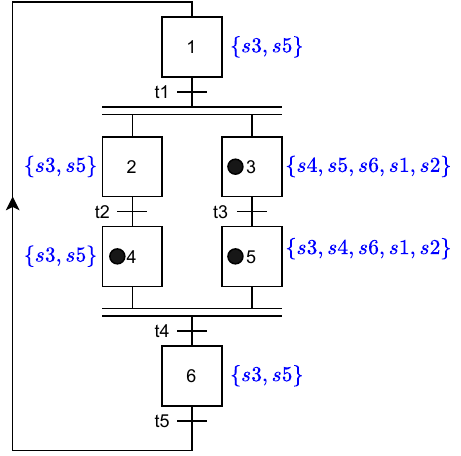}
	\caption{Example partial Grafcet $c$ with the initial situation $\{s3, s4, s5\}$ and the possible concurrent steps $S_s^C$ marked in blue for every step $s \in S_c$, which is the final result of Alg. \ref{alg:reachabilility} and \ref{alg:concurrency}.}
	\label{fig:exampleconcurrencyalg}
\end{figure}
\begin{figure*}[t]
	\centering
	\includegraphics[width=.8\textwidth]{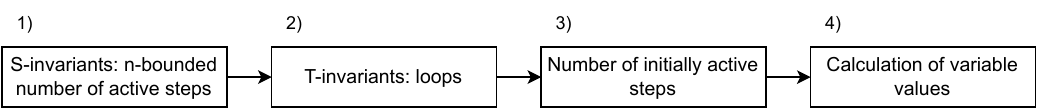}
	\caption{Overview of approximation of internal and output variables.}
	\label{fig:structuralAnalysis}
\end{figure*}

To analyze concurrent steps, we extend Alg.~\ref{alg:reachabilility} by Alg.~\ref{alg:concurrency}. It calculates for every step $s \in S_c$ a set $S^C_s$ of steps concurrent to $s$ called $\mathcal{S} = \{S_s^C\}_{s \in S_c}$. The basic assumption is that the downstream steps $t \bullet$ of a transition $t$ have the same concurrent steps as the upstream steps $\bullet t$. This is true for basic sequences without activation of parallel sequences like e.g., in G6 in Fig.~\ref{fig:concurrentStructures}.  
When a parallel sequence is activated (i.e., $|t\!\bullet\!| > 1$) all downstream steps $s \in t \bullet$ become concurrent to each other (line 2). With synchronizations (i.e., $|\!\bullet\! t| > 1$) the opposite is the case. Therefore, in line 3 we intersect the concurrent steps $S^C_{s'}$ of the upstream steps $s' \in \bullet t$ of $t$. This propagates the concurrent steps along the transitions except for synchronized steps. 
Since concurrency is a symmetric relation, the algorithm adds the current step $s$ to all its concurrent steps $s''$ in lines 4 - 5.

To initialize the concurrency analysis, we set all initially active steps concurrent to each other ($S_s^C \leftarrow S^I \backslash \{s\}$ if $s \in S^I$, $\emptyset$ otherwise).

Source transitions are a special case since they can activate the downstream steps concurrently to all other active steps in every possible situation. To model this behavior we calculate $S^R$ as presented above and run Alg. \ref{alg:reachabilility} again with an initial $\mathcal{S}$ that represents this behavior (i.e., $S_s^C \triangleleft S^R$ if $s$ is a downstream step of a source transition).

To illustrate the algorithm, cf. Fig. \ref{fig:exampleconcurrencyalg}, with an initial situation $S^I = \{s3, s4, s5\}$ indicated by the black dots induced, e.g., by a forcing order. Therefore, we initialize $W \leftarrow \{t3, t4\}$ and $\mathcal{S} \leftarrow \{S_{s1}^C, \dotsb, S_{s6}^C \}$, where $S_{s3}^C = \{s4, s5\}, S_{s4}^C = \{s3, s5\}, S_{s5}^C = \{s3, s4\}$ and $S_{s1}^C = S_{s2}^C = S_{s6}^C = \emptyset$. Assuming we withdraw $t = t4$ from the worklist in the first iteration of Alg. \ref{alg:reachabilility}, we mark $s = s6$ as reachable in line 8. Executing Alg. \ref{alg:concurrency} we have $\mathbin{t \mathbin{\bullet}} \backslash \{s\} = \emptyset$ and  $\bigcap_{s' \in \bullet t} S_{s'}^C = \{s3, s5\} \cap \{s3, s4\} = \{s3\}$. In line 5 of Alg.~\ref{alg:concurrency} we add $s6$ to $S_{s3}^C$. $t5$ (due to line 9 in Alg. \ref{alg:reachabilility}) and $t3$ (due to line 13 in Alg. \ref{alg:reachabilility}) are added to the worklist and the algorithm iterates along the transitions as long as $\mathcal{S}$ does change. 
This results in values for $\mathcal{S}$ indicated in Fig. \ref{fig:exampleconcurrencyalg} by the sets next to the steps.


\subsection{Approximation of internal variables and conditions} 
\label{ssec:flowinsens}

In the previous Sec.~\ref{ssec:stepApprox}, we proposed an analysis to approximate the step variables of GRAFCET. 
In this section, we propose an analysis to approximate the internal and output variables written by stored actions. 
Fig. \ref{fig:structuralAnalysis} shows an overview of the proposed analysis. To evaluate the variable values, we over-approximate the action's number of executions and assume the order to be non-deterministic. The number of executions is influenced by multiple structural elements of GRAFCET. Those are: 1) source transitions (Fig. \ref{fig:concurrentStructures}, G5) and the equivalent with an activation of parallel sequences (Fig. \ref{fig:concurrentStructures}, G4), 2) loops as well as 3) multiple active steps in a sequence (Fig. \ref{fig:concurrentStructures}, G2). The influence of the different structural elements on how often an action might be executed is covered by the respective steps 1) - 3) of the analysis (cf. Fig.~\ref{fig:structuralAnalysis}).

First, we calculate the S-invariants of the Grafcet. 
If the Grafcet is covered with S-invariants (i.e., for every step there is at least one S-invariant with a value $n \in \mathbb{N}$ for the corresponding step) it is a sufficient condition that the number of how often a step can be activated is bounded by $n$ (without taking loops and multiple initially active steps into account, which are covered in steps 2) and 3)). 
Otherwise the possible number of the steps' activation is considered to be infinite and therefore, all actions not covered with S-invariants (i.e., there is no S-invariant with a value $n \not= 0$ for the corresponding step $s_a$ of the action $a$) can be executed infinitely often. 
This detects structures like in G4 (with a S-invariant $\mathbf{y}_{G4} = (1, 0, 0)$ that does not cover steps $2$ and $3$) and G5 (with no existing S-invariant) in Fig. \ref{fig:concurrentStructures}. On the other hand structures like in G6 are covered with S-invariants ($\mathbf{y}_{G6, 1} = (1, 1, 0, 1, 0)$ and $\mathbf{y}_{G6, 2} = (1, 0, 1, 0, 1)$).
However, even if a Grafcet is $n$-bounded, loops can cause actions to execute infinitely often. Therefore, in the $n$-bounded case we calculate the T-invariants (step 2) in Fig. \ref{fig:structuralAnalysis}) to detect loops to approximate the number of executions further (e.g. the structure in G7, Fig. \ref{fig:concurrentStructures} has a T-invariant $\mathbf{x}_{G7} = (1, 1)$). In case of a loop we assume the associated actions to be executed infinitely often. In step 3) in Fig. \ref{fig:structuralAnalysis} we consider the number of initially active steps $S^I$ that can cause a multiple executions of actions as well. 
If  the actions' number of executions is not infinite it is calculated with $n\cdot |S^I|$.
\begin{figure}
	\centering
	\includegraphics[width=0.45\linewidth]{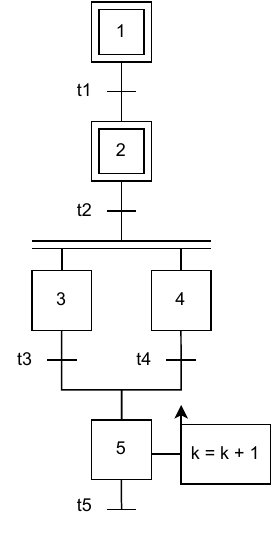}
	\caption{Example partial Grafcet with $S^I = \{s1, s2\}$ that is $n$-bounded with $n = 2$.}
	\label{fig:exampleinvariants}
\end{figure} 
The approach can be illustrated using the partial Grafcet in Fig. \ref{fig:exampleinvariants}. It has one S-invariant $\mathbf{y} = (2, 2, 1, 1, 1)$ and is therefore $n$-bounded with $n = 2$. The partial Grafcet has no T-invariant and therefore, no loop. Due to the two initial steps $1$ and $2$ in combination with the Grafcet being $2$-bounded, step $5$ can be activated four times since $n \cdot |S^I| = 2 \cdot 2 = 4$. Note that a value of $n > 1$ could be considered a design error by itself. 

To calculate the input and output variable values (step 4) in Fig. \ref{fig:structuralAnalysis}) and to take into account the non-deterministic order of the action's execution, we select the order resulting in the largest result to guarantee over-approximation and therefore soundness.
We calculate all possible values $X_{\mathit{calc}}$ for every variable using the number of executions of the corresponding actions identified in the first three steps. Then we choose the minimal and maximal result as lower and upper bound of the variable value's interval or the initialization value zero to approximate the variable value: $[\mathrm{min}(\mathrm{min}(X_{\mathit{calc}}), 0), \mathrm{max}(0, \mathrm{max}(X_{\mathit{calc}}))]$. For the example in Fig. \ref{fig:exampleinvariants} the action on activation associated to step $5$ can be executed four times as shown above. The resulting interval for $k$ evaluates to $[0, 4]$. 

In the end we can check if the conditions in transitions and actions are satisfiable using the approximated internal and step variables.

%% file: p_evaluation.tex
\section{Evaluation}
\label{sec:eval}
\begin{figure*}
	\centering
	\includegraphics[width=0.7\linewidth]{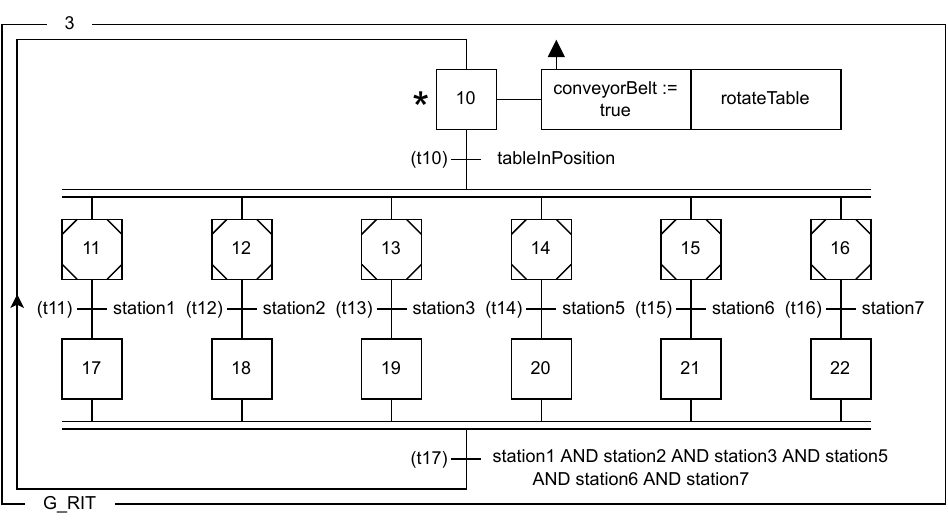}
	\caption{GRAFCET specification G\_RIT of the control of the rotary indexing table as well as the stations.}
	\label{fig:g_rit}
\end{figure*}
\begin{figure}
	\centering
	\includegraphics[width=0.85\linewidth]{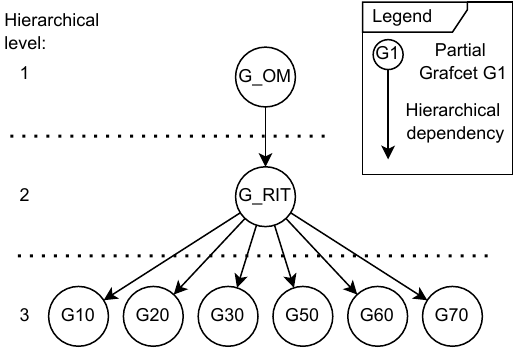}
	\caption{Hierarchical dependencies of the testing machine's GRAFCET specification.}
	\label{fig:hierarchiegraphschumacher}
\end{figure}

The proposed analysis was implemented and integrated in a toolchain developed by the authors. Part of the toolchain is a graphical editor for GRAFCET based on a GRAFCET meta-model proposed by Julius et al. \cite{Julius.19}. The meta-model was implemented using the Eclipse Modeling Framework (EMF)\footnote{\url{https://www.eclipse.org/modeling/emf/}}.

The presented analysis was evaluated using the GRAFCET specification of an industrial plant first shown in \cite{Schumacher.14}. 
The application example is an automatic testing machine for quality control of components and consists of a conveyor belt, a rotary indexing table and six stations. Coordinated by the rotary indexing table, the parts pass through these stations, where separation and quality control take place. The components are marked as regular or damaged parts, and damaged parts are subsequently sorted out.
The complete specification consists of eight partial Grafcets shown in Fig. \ref{fig:hierarchiegraphschumacher} including their hierarchical dependencies. On the top level the partial Grafcet G\_OM determines the operation mode. 
Choosing the automatic operation mode, G\_RIT is activated and controls the rotary indexing table as well as the six stations represented by G10 to G70. 
All hierarchical dependencies are induced by enclosing steps which are used to implement an emergency stop. 
G\_OM is activated by an initial step. 
Altogether the specification consists of 60 steps, 62 transitions 
and 80 Boolean and integer variables. 

Fig.~\ref{fig:g_rit} shows the partial Grafcet G\_RIT\footnote{The full specification formalized with GRAFCET can be viewed here: \url{https://github.com/Project-AGRAFE/GRAFCET-instances}}. The number 3 at the top refers to the enclosing step 3 in G\_OM controlling G\_RIT. The asterisk at step 10 marks the step activated by the enclosing step.
The station activates the conveyor belt by setting the output variable \textit{conveyorBelt} to true and rotates the rotary indexing table by one sector using the output variable \textit{rotateTable}. This transfers the parts from one station to another and can therefore only occur while the stations are in their starting position. The activation of the stations happens in parallel via enclosing steps 11 to 16. The internal Boolean variables \textit{station1}, \textit{station2}, etc. indicate if a station is finished. Since these variables are read in G\_RIT and written in G10 to G70 they are affected by concurrency. 

Applying the analysis regarding the structural reachability and concurrency from Sec.~\ref{ssec:stepApprox} to G\_RIT results in all steps being reachable, as well as the following concurrent steps: step~10 has no concurrent steps - step~11 and 17 are structurally concurrent to the steps~12, 13, 14, 15, 16, 18, 19, 20, 21, 22 - the concurrent steps for the remaining steps are analog to step~11 and 17. We applied the analysis to the entire specification, taking into account that the enclosing steps of G\_RIT are concurrent to each other, and therefore the stations run concurrently as well. The execution time was less than ten milliseconds. We used this information to confirm that no race conditions are present in the Grafcet, i.e., no stored actions writing the same variable are associated to concurrent steps. 

By applying the analysis from Sec.~\ref{ssec:flowinsens} shown in Fig.~\ref{fig:structuralAnalysis} to G\_RIT, we start by calculating the S-invariants $\mathbf{y}_1 = (s10, s11, s17)$,  $\mathbf{y}_2 = (s10, s12, s18)$, $\mathbf{y}_3 = (s10, s13, s19)$, etc., which cover the whole Grafcet (note that we misuse the notation here by writing the step name when we mean that there is a corresponding 1 and omit it when there is a 0). 
Thus, the number of active steps for every S-invariant is 1-bounded and no structures like shown in Fig. \ref{fig:concurrentStructures}, G4 are present. In the next step the T-invariants are calculated resulting in one invariant $\mathbf{x}_1 = (t10, t11, t12, t13, t14, t15, t16, t17)$, which means that all covered transitions (i.e., all transitions that have a corresponding value of 1 in $\mathbf{x}_1$) have to fire for the initial situation to be reached again.
Therefore, the actions associated to step 10 can be executed infinitely often resulting in a possible value assignment \textit{conveyorBelt} $=$ \textit{rotateTable} $=$ \{\textit{false}, \textit{true}\}. Applying the analysis to all other partial Grafcets results in the same possible value set for \textit{station1} to \textit{station7} and therefore, \textit{t11} to \textit{t17} can fire according to the analysis, which is the expected behavior. 

This allows to identify safety critical situations. E.g., in the given example, the rotary indexing table must not rotate (i.e., \textit{rotateTable} $=$ \textit{false}) as long as one of the six stations is working, 
or vice versa. Using the approximation of the variable values, the analysis returns a false alarm for this requirement being violated since it can only detect if the variable values will eventually be true or false, but not if this will be at the same time or sequentially. However, using the results from the analysis proposed in Sec.~\ref{ssec:stepApprox}, it can be shown that the stations are deactivated while the rotary indexing table is rotating: As shown in Fig.~\ref{fig:g_rit}, the variable \textit{rotateTable} is \textit{true} when step~10 is active and since step~10 has no concurrent steps the stations can not be active at the same time.
 

%% file: p_conclusion.tex
\section{Conclusion}
\label{sec:conclusion}
In Sec.~\ref{sec:contribution}, we demonstrated that, owing to GRAFCET's concurrent behavior, control flow based analysis approaches from the field of sequential programs are inapplicable to GRAFCET. Further, we determined which GRAFCET structures and elements cause concurrent behavior.
The resulting approach resolves this challenge by over-approximation and neglecting transition conditions. 
We presented a worklist algorithm to approximate the step variables as well as their concurrency. Further we presented how analysis means from the field of Petri nets can be used to approximate internal and output variable values. 
Therefore, we demonstrated an approach that uses a structural analysis to verify possible GRAFCET instances including elements proposed by the standard IEC 60848 that can result in concurrent behavior. Despite the resulting over-approximation (which can lead to a report of concurrent steps where no such behavior is present), we presented in the evaluation that the approach provides valuable information, such as verifying that no writing conflicts exist. 

For future work we want to reduce the degree of over-approximation. One approach is to analyze the GRAFCET instances depending on whether concurrency is present or not. It can be assumed that not all Grafcets show all different elements that result in concurrent behavior. Instead of an analysis that can handle all structures proposed by the standard it might be beneficial to provide specialized algorithms that can analyze only a subset of the possible GRAFCET instances. Depending on the Grafcet to be analyzed, a specialized analysis that might result in less over-approximation could be chosen. 
Another approach for future work would be to use the information obtained from the concurrency analysis to track the interference between concurrent steps induced by actions writing variable values (e.g., when a variable is incremented due to the activation of a concurrent step). This interference could be used to analyze the Grafcets by means of abstract interpretation, similarly to works, e.g., proposed by Min{\'e} \cite{Mine.14}.

%% file: ETFA_23.bbl
\begin{thebibliography}{10}
\providecommand{\url}[1]{#1}
\csname url@samestyle\endcsname
\providecommand{\newblock}{\relax}
\providecommand{\bibinfo}[2]{#2}
\providecommand{\BIBentrySTDinterwordspacing}{\spaceskip=0pt\relax}
\providecommand{\BIBentryALTinterwordstretchfactor}{4}
\providecommand{\BIBentryALTinterwordspacing}{\spaceskip=\fontdimen2\font plus
\BIBentryALTinterwordstretchfactor\fontdimen3\font minus
  \fontdimen4\font\relax}
\providecommand{\BIBforeignlanguage}[2]{{%
\expandafter\ifx\csname l@#1\endcsname\relax
\typeout{** WARNING: IEEEtran.bst: No hyphenation pattern has been}%
\typeout{** loaded for the language `#1'. Using the pattern for}%
\typeout{** the default language instead.}%
\else
\language=\csname l@#1\endcsname
\fi
#2}}
\providecommand{\BIBdecl}{\relax}
\BIBdecl

\bibitem{Schumacher.13b}
F.~Schumacher, S.~Schröck, and A.~Fay, ``{Tool support for an automatic
  transformation of GRAFCET specifications into IEC 61131-3 control code},'' in
  \emph{2013 IEEE 18th Conference on Emerging Technologies \& Factory
  Automation (ETFA)}, 2013, pp. 1--4.

\bibitem{Julius.19}
R.~Julius, T.~Trenner, A.~Fay, J.~Neidig, and X.~L. Hoang, ``{A meta-model
  based environment for GRAFCET specifications},'' in \emph{2019 IEEE
  International Systems Conference (SysCon)}, 2019, pp. 1--7.

\bibitem{iec60848}
{IEC 60848}, ``{GRAFCET specification language for sequential function
  charts},'' International Electrotechnical Commission, IEC 60848, 2013.

\bibitem{Provost.11}
J.~Provost, J.-M. Roussel, and J.-M. Faure, ``{A formal semantics for Grafcet
  specifications},'' in \emph{2011 IEEE International Conference on Automation
  Science and Engineering}, 2011, pp. 488--494.

\bibitem{VogelHeuser.14}
B.~Vogel-Heuser, C.~Diedrich, A.~Fay, S.~Jeschke, S.~Kowalewski,
  M.~Wollschlaeger, and P.~Göhner, ``{Challenges for Software Engineering in
  Automation},'' \emph{Journal of Software Engineering and Applications},
  vol.~7, pp. 440--451, 2014.

\bibitem{Mross.22}
R.~Mross, A.~Schnakenbeck, M.~Völker, A.~Fay, and S.~Kowalewski,
  ``{Transformation of GRAFCET Into GAL for Verification Purposes Based on a
  Detailed Meta-Model},'' \emph{IEEE Access}, vol.~10, pp. 125\,652--125\,665,
  2022.

\bibitem{Julius.17}
R.~Julius, M.~Sch{\"u}renberg, F.~Schumacher, and A.~Fay, ``{Transformation of
  GRAFCET to PLC code including hierarchical structures},'' \emph{Control
  Engineering Practice}, vol.~64, pp. 173--194, 2017.

\bibitem{Boehm.1981}
B.~W. Boehm, \emph{{Software engineering economics}}, ser. Prentice-Hall
  advances in computing science and technology series.\hskip 1em plus 0.5em
  minus 0.4em\relax Englewood Cliffs, NJ: Prentice-Hall, 1981.

\bibitem{Cabasino.13}
M.~P. Cabasino, A.~Giua, and C.~Seatzu, \emph{Structural Analysis of Petri
  Nets}.\hskip 1em plus 0.5em minus 0.4em\relax London: Springer London, 2013,
  pp. 213--233.

\bibitem{Praehofer.12}
H.~Prähofer, F.~Angerer, R.~Ramler, H.~Lacheiner, and F.~Grillenberger,
  ``{Opportunities and challenges of static code analysis of IEC 61131-3
  programs},'' in \emph{2012 IEEE 17th International Conference on Emerging
  Technologies \& Factory Automation (ETFA)}, 2012, pp. 1--8.

\bibitem{Sogbohossou.20}
\BIBentryALTinterwordspacing
M.~Sogbohossou and A.~Vianou, ``{Translation of hierarchical GRAFCET charts
  into time Petri nets},'' Sep. 2020, working paper or preprint. [Online].
  Available: \url{https://hal.archives-ouvertes.fr/hal-02934113}
\BIBentrySTDinterwordspacing

\bibitem{Lesage.93}
\BIBentryALTinterwordspacing
J.-J. Lesage and J.-M. Roussel, ``{Hierarchical approach to GRAFCET using
  forcing order},'' \emph{{Automatique Productique Informatique Industrielle}},
  vol.~27, no.~1, pp. 25--38, Mar. 1993. [Online]. Available:
  \url{https://hal.archives-ouvertes.fr/hal-00347044}
\BIBentrySTDinterwordspacing

\bibitem{Lesage.96}
J.-J. Lesage, J.-M. Roussel, and C.~Thierry, ``{A theory of binary signal},''
  \emph{Proceedings of the IEEE multiconference on computational engineering in
  systems applications}, p. 590–595, 1996.

\bibitem{Schumacher.13a}
F.~Schumacher and A.~Fay, ``{Transforming time constraints of a GRAFCET graph
  into a suitable Petri net formalism},'' \emph{2013 IEEE International
  Conference on Industrial Technology (ICIT)}, pp. 210--218, 2013.

\bibitem{Schumacher.14}
------, ``{Formal representation of GRAFCET to automatically generate control
  code},'' \emph{Control Engineering Practice}, vol.~33, pp. 84--93, 2014.

\bibitem{Bonet.07}
P.~Bonet and C.~M. Llad{\'o}, ``Pipe v2.5 : A petri net tool for performance
  modeling,'' \emph{Proc. 23rd Latin American Conference on Informatics (CLEI
  2007), San Jose, Costa Rica, Oct.}, 2007.

\bibitem{Girault.03}
C.~Girault and R.~Valk, \emph{{Petri Nets for Systems Engineering}},
  1st~ed.\hskip 1em plus 0.5em minus 0.4em\relax Heidelberg, Germany: Springer
  Berlin, 2003.

\bibitem{Mine.14}
A.~Min{\'e}, ``Relational thread-modular static value analysis by abstract
  interpretation,'' in \emph{Verification, Model Checking, and Abstract
  Interpretation}, K.~L. McMillan and X.~Rival, Eds.\hskip 1em plus 0.5em minus
  0.4em\relax Berlin, Heidelberg: Springer Berlin Heidelberg, 2014, pp. 39--58.

\end{thebibliography}
